\documentclass[review]{elsarticle}

\journal{Computer Communications}

\usepackage{graphicx}
\usepackage{multirow}
\usepackage{float}
\usepackage{subcaption}
\usepackage{caption}
\usepackage{subcaption}
\usepackage{amsthm}
\usepackage{hyperref}
\usepackage{amssymb}
\usepackage{amsfonts}
\usepackage{amsmath}
\usepackage{hyphenat}

\usepackage{hyperref}
\usepackage{sty/lineno/lineno}

\graphicspath{{Figures/}}
\begin{document}
\begin{frontmatter}

\title{Large Scale Model for Information Dissemination with Device to Device Communication using Call Details Records}

%
%

\author[mainaddress]{Rachit Agarwal}
\author[mainaddress]{Vincent Gauthier \corref{correspondingauthor}}
\author[mainaddress]{Monique Becker}
\author[mainaddress,secondaryaddress]{Thouraya Toukabrigunes}
\author[mainaddress]{Hossam Afifi}
\cortext[correspondingauthor]{Corresponding author}
\cortext[email]{Email addresses of the authors are \{rachit.agarwal, vincent.gauthier, monique.becker, hossam.afifi\}@telecom-sudparis.eu and thouraya.toukabrigunes@orange.com}
\address[mainaddress]{Lab. CNRS SAMOVAR UMR 5157, Telecom SudParis/Institut Mines-Telecom.}
\address[secondaryaddress]{Orange Lab.}

%
%
\begin{abstract}
In a network of devices in close proximity such as \emph{Device to Device} ($D2D$) communication network, we study the dissemination of public safety information at country scale. In order to provide a realistic model for the information dissemination, we extract a spatial distribution of the population of Ivory Coast from census data and determine migration pattern from the Call Detail Records ($CDR$) obtained during the \emph{Data for Development} ($D4D$) challenge. We later apply epidemic model towards the information dissemination process based on the spatial properties of the user mobility extracted from the provided $CDR$. We then propose enhancements by adding latent states to the epidemic model in order to model more realistic user dynamics. Finally, we study dynamics of the evolution of the information spreading through the population.
\end{abstract}

\begin{keyword}
Device to Device communication, Human Mobility, Information Dissemination, Opportunistic networks.
\end{keyword}

\end{frontmatter}
%
%
\section{Introduction}\label{sec:introduction}

In communication networks, devices in each other's physical proximity can communicate. Information dissemination in such proximity based networks have been the focus of a lot of studies. Through this paper, we study the dissemination of public safety information at country scale. For the type of communication network, as a case study, we choose \emph{Device to Device} ($D2D$) communication paradigms. This choice relates to the fact that the market of context-aware applications and location-based services has grown tremendously and operators have started to consider the deployment of $D2D$ communications as an underlay to the cellular networks. $D2D$ communication is defined as a short range communication between devices in physical proximity without any involvement of the network infrastructure. $D2D$ has many advantages like, autonomous communication, improved performance and spectrum reuse, low energy consumption and reduced load on the infrastructure. Moreover, other benefits of $D2D$ communication include direct communication between devices even when the traditional infrastructure is down, better connectivity in poor connectivity regions and increased average rate of successful message delivery. These above mentioned benefits of $D2D$ motivate us to choose $D2D$ communication paradigms as a Use Case. Nevertheless, our Use Case can also be applied to mobiles applications such as \texttt{Firechat} \cite{firechat} that leverages direct connection between devices in close proximity to temporally connect users.

Moreover, in opportunistic networks, lot of studies have focus on the store-carry-and-forward paradigm and on the dynamics of the message delivery. The first attempt to model the process was represented via epidemic forwarding. Variations of the process includes the integration of \emph{Social Network Analysis} measures to study the dissemination process \cite{5677535, 4674358, Mtibaa2013180, Belblidia20121786}. Further, modeling network performance through Markov Chains in opportunistic networks has also been the focus of several studies \cite{Groenevelt2005210,AlHanbali2008463,4430783, Boldrini201456}. However, none of the methods are scalable to the country scale since they rely on computing individual transmission probability. Dissemination methods based on mean field approach, however, are more scalable \cite{Zhang:2007:PME:1242848.1243153,1597221} but rely upon homogeneous contact dynamics that is far from being valid. Thus, recent studies \cite{colizzaepidemic2008, balcanphase2011, polettoheterogeneous2012, wanghow2013} on dissemination process are based on heterogeneous flow within the \emph{metapopulation}. This  enabling us to study and create a heterogeneous population model. These studies offer new perspectives on such models and enable us to study more easily country wide dissemination process.

Since, our interest lies in studying the dissemination process on a countrywide scale, we consider the population of the device to be spatially structured into subpopulation of well mixed individuals where the diffusion of safety information takes place. We use the official boundaries (termed as \emph{subprefecture}) in \emph{Ivory Coast} to define the subpopulation structure of our \emph{metapopulation} model. In order to know the flow of people between the subprefectures, we use the Call Detail Records ($CDR$) provided by the Orange Labs during the Data for Development ($D4D$) Challenge \cite{Blondel2012}. For future reference, we refer the provided $CDR$ as the $D4D$ dataset. The analysis of the movement pattern of the people in Ivory Coast reveals the flow of people between subprefectures. $D4D$ dataset enable us to monitor trajectories of people at unprecedented scale and derive key spatio-temporal patterns of the movement. Thus, the mobility is modeled as a network of interactions between each subpopulation (population within a subprefecture) where the connections correspond to the flow of people among them. Metapopulation\cite{Arino2006, Watts2005} model has been studied towards diffusion of information combined with human mobility where mobility pattern show strong heterogeneity \cite{gonzálezunderstanding2008, brockmannthe2006, colizzaepidemic2008}. Moreover, the outcome of the diffusion of the public safety information is regulated by the coupling of the mobility process and the diffusion process in each subpopulation. Thus, it is of utmost importance that the mobility model correctly describes realistic pattern of mobility.

However, a simple memoryless mobility model is far from being able to model recurring patterns such as commuting behavior or traveling behavior that are often observed in mobility data \cite{gonzálezunderstanding2008, songlimits2010}. To include this effect we use a non-markovian diffusive process to cope with the recurring behavior \cite{balcanphase2011, polettoheterogeneous2012}. In a realistic human population, a user is associated with a home location. Every time a user moves away from the home location, there is a probability to return to the home location from the new location. This return probability is also derived from the $D4D$ dataset. More details on how the return probability is computed is presented in the section \ref{subsec:Data set_2_and_3}.

The $D4D$ dataset includes the information about the user mobility and the association of a user to a community, termed as subprefecture. Further, in order to study the model with realistic population structure, we use the census data provided in \cite{Web1} to populate each subprefecture (cf. Section \ref{subsec:Data set_2_and_3}).

Thus, in this paper, we use insights from the dissemination in the metapopulation model and the $D4D$ dataset to formulate our analytical model. Further, our model is divided into various stages. First, construct synthetic yet realistic population of the devices using census data \cite{Web1} for the region of Ivory Coast. Then extract mobility information from the $D4D$ dataset. Then find mobility steady state population in each community. Finally, apply Susceptible($S$)-Infected($I$)-Recovered($R$) type epidemic modeling (cf. Section \ref{subsec:C3_dissmodel}) to model the information dissemination process at the countrywide scale.

In order to reduce the complexity of the dissemination process at country scale, we use the reactive-diffusive equation with homogeneous assumption to describe the subpopulation at the mechanistic level. We use a variation of the $SIR$ epidemic model where Susceptible ($S$), Infected ($I$) and Recovered ($R$) states correspond to the state of a device `not having the message', `having the message and transmitting it' and `having received the message but stop transmitting it due to message timeout (message time to live, $TTL$)', respectively. A comprehensive survey about the epidemic model is provided in \cite{Hethcote2000,Britton2010}. Recently, the authors of \cite{Lund2013} showed that not only the community structure affects the dissemination process but the density of the communities also play an important role in the dissemination process. In metapopulation model it is often assumed that the population within a single subpopulation is well mixed. This is of course not generally true specially in a large subpopulation. To acknowledge this in a large subpopulation where some devices may never be in contact at a certain time, we include Latent states ($E$) in our model. Device in latent state account for devices that belong to a given subpopulation but are not participating in the diffusion process. To distinguish between devices that are Latent but Susceptible, Latent but Infected and Latent but Recovered, we subdivide $E$ into three states: $E_{S}$, $E_{I}$ and $E_{R}$. In our model only active devices in $S$, $I$ and $R$ states participate in the dissemination process. Thus, the novelty of the paper lies in the introduction of latent states and in the way we compute the number of interactions per device.

The remainder of this paper is organized as follows. In Section \ref{sec:D2Dscenario} we first provide an overview of the $D2D$ scenario. In Section \ref{sec:C3_Data Analysis} we describe how the dataset provided by the $D4D$ organizers is used to get useful information. A detailed description of the model is then provided in Section \ref{sec:model} which is followed by the results obtained in Section \ref{sec:simresult}. The paper finally concludes with Section \ref{sec:conclusion}.

%
%

\section{Loosely controlled $D2D$ support for information dissemination}\label{sec:D2Dscenario}
\begin{figure}
    \centering
    \includegraphics[width=0.6\textwidth]{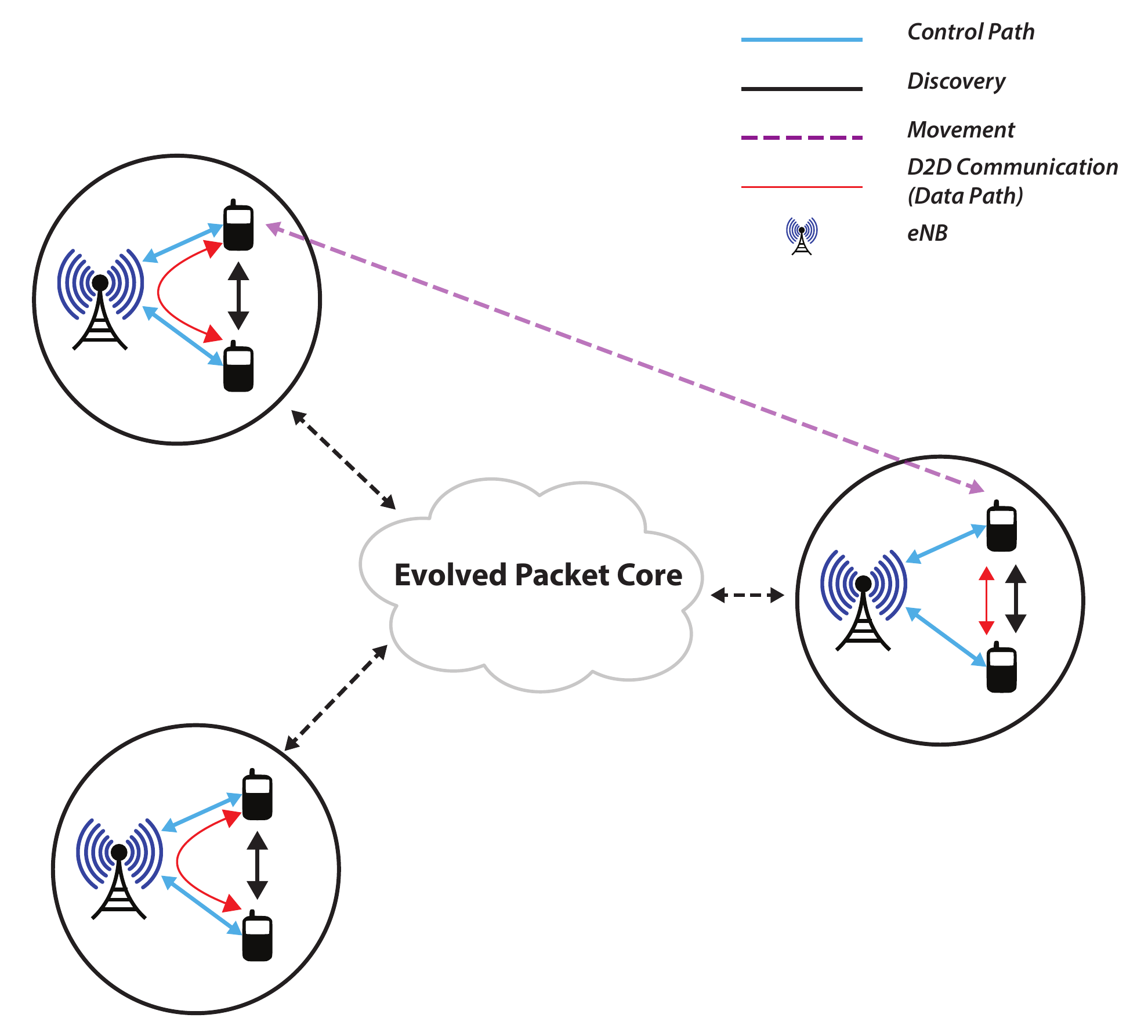}
    \caption{$D2D$ communication scheme.}
    \label{fig:d2d}
\end{figure}
We assume the following $D2D$ scenario: Using a Public Warning System (PWS), an $eNB$ (evolved Node B, a radio interface in the Long Term Evolution (LTE) network) has to broadcast a national emergency warning which should reach the maximum number of people in the country. However, in such situations, the operator's network is massively overloaded by a huge traffic of users trying to use their mobile devices at the same time. Also, the warning message may not reach a part of the population due to network coverage issues in rural areas.
Using $D2D$ communications, devices in the physical proximity can communicate directly with each other and exchange messages. This mechanism allows not only an extension to the broadcasting area but also enhances the traffic on the network infrastructure and avoids network saturation and waste of resources. A $D2D$ communication has two phases: the Neighbor Discovery phase and the Communication phase for the data exchange. Both phases can be based on either a direct approach or a network-assisted approach \cite{Fodor2102,Lei2012,3GPP13,3GPP14,Toukabri14}.

In the direct $D2D$ approach, devices discover other devices in their surrounding by exchanging presence beacons. Both the phases, the discovery and the communication phase, are done in an adhoc-like way without any assistance or control of the network infrastructure. The communicating devices then form a self-organized and a self-configurable network. This approach is flexible and highly scalable as it can adapt to an increasing number of connected devices and active $D2D$ links. This allows offloading the traffic of the core operator network to local $D2D$ communication. Besides, it is also a solution to avoid communication disruption when some $eNBs$ are down in a disaster situation like Earthquake or when an $eNB$ fails. The users could still use their devices and be connected to the operator network using direct communication.

In the centralized $D2D$ approach, the operator network may fully control the discovery and communication phases (fully controlled $D2D$) or assist the devices during the whole $D2D$ process (loosely controlled $D2D$) by enabling authentication and QoS mechanisms. Nevertheless, this approach is less scalable than the direct approach as it consists of performance and load balancing issues at the Radio Access Network (RAN) when dealing with a huge number of $D2D$ connections. As depicted in Fig. \ref{fig:d2d} we shows that some $eNBs$ can allow the direct communications between devices that are in physical proximity while some do not.

In addition, a device can move from one $eNB$ to another. Moreover, the device can also move within an associated $eNB$. In this paper, we assume that all $eNBs$ allow direct communication between devices that are in physical proximity and that a public warning message is generated in some area associated to an $eNB$. In the section \ref{sec:model}, we are interested in investigating how fast dissemination of the warning message can be achieved through a large population in a $D2D$ environment where the devices are also mobile and links are intermittent. By fast dissemination we mean how, after the warning message is generated, a large set of devices can quickly receive the warning message.

%
%
\section{Data}\label{sec:C3_Data Analysis}
\captionsetup[subfigure]{labelformat=empty}
\begin{figure*}
    \centering
    \begin{subfigure}[b]{0.35\textwidth}
         \includegraphics[width=\textwidth]{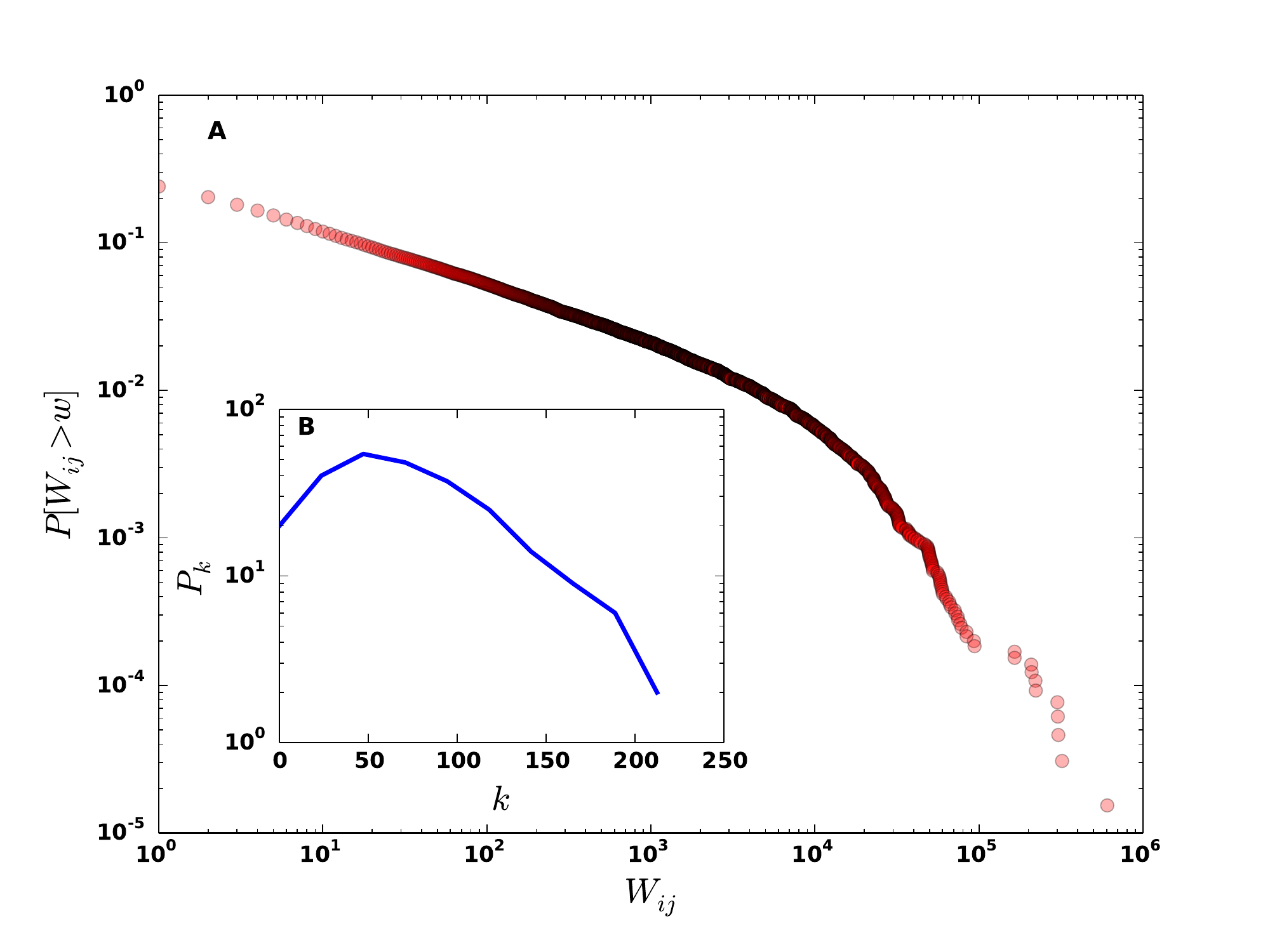}
	\caption{}
	\label{fig:distributiona}
    \end{subfigure}	
    \begin{subfigure}[b]{0.35\textwidth}
         \includegraphics[width=\textwidth]{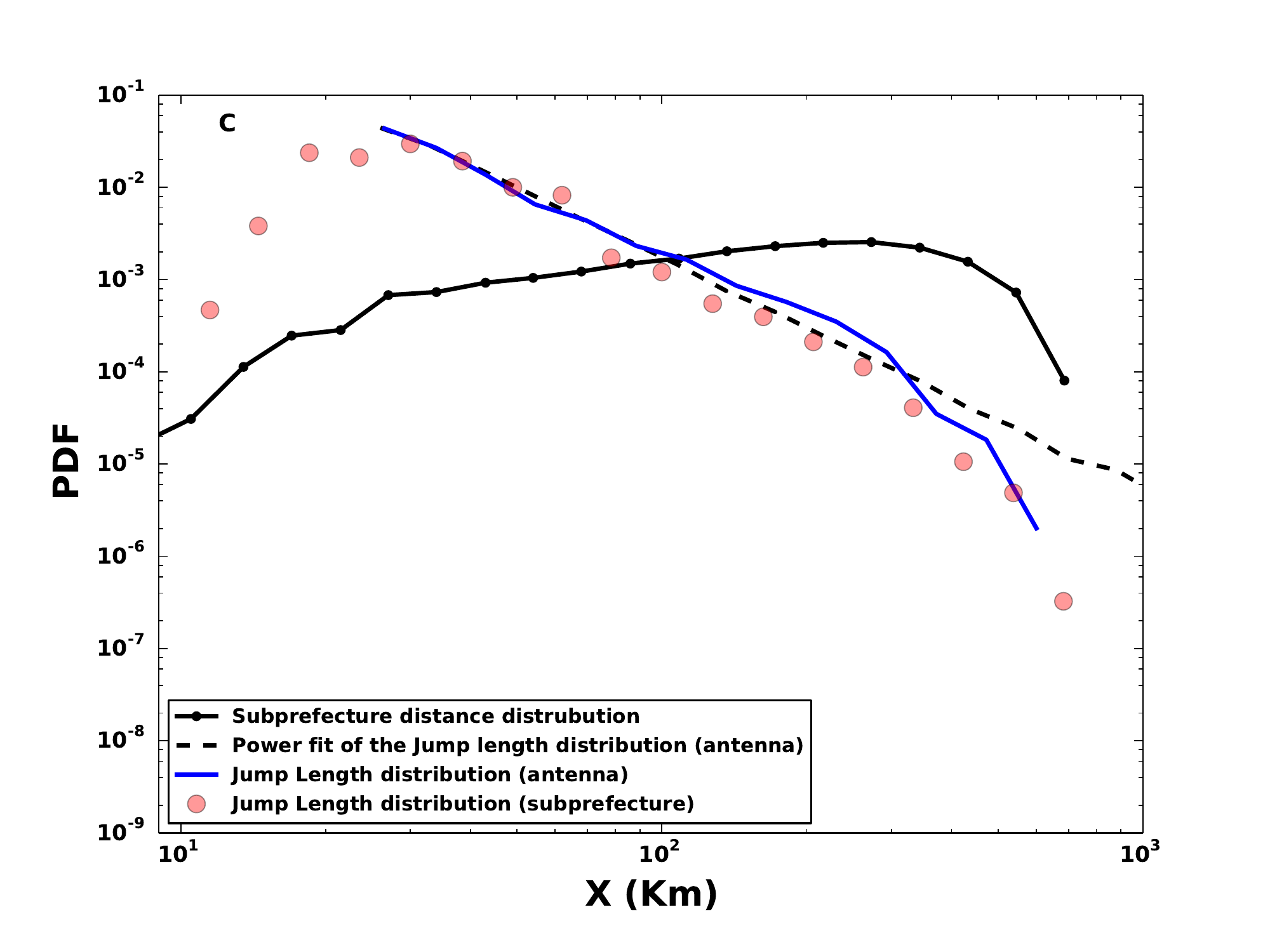}
         \caption{}
	\label{fig:distributionb}
    \end{subfigure}
    \begin{subfigure}[b]{0.28\textwidth}
         \includegraphics[width=\textwidth]{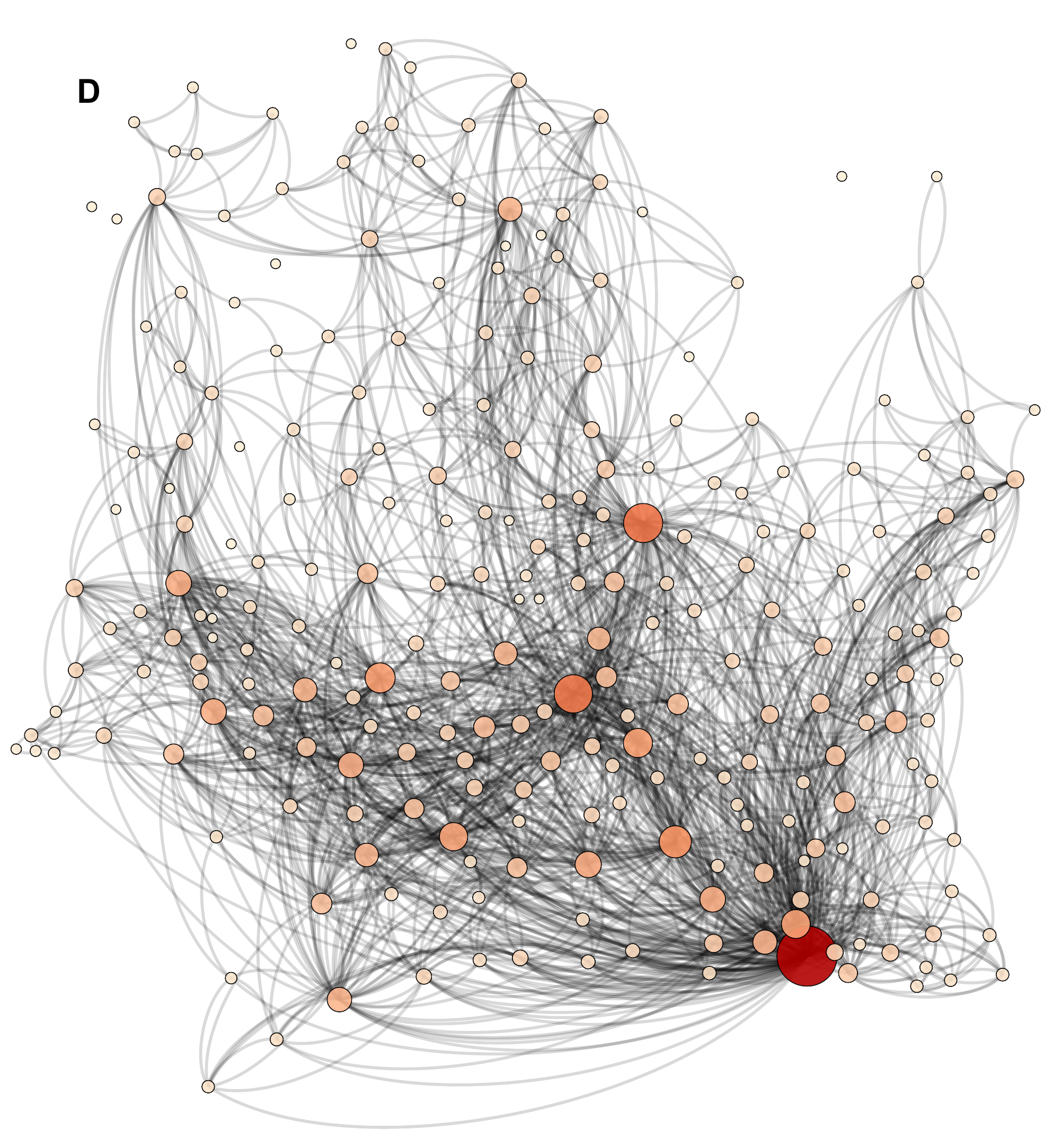}
         \caption{}
	\label{fig:graph}
    \end{subfigure}
    \caption{\textbf{a)} CCDF of flow distribution of the users obtained by aggregating all the transition made by user in the SET3 \textbf{b)} the degree distribution of the transition graph \textbf{c)} Transition graph \textbf{d)}}
    \label{fig:distribution}
\end{figure*}

In this Section we provide detailed description of the $D4D$ dataset that we use and the methods we apply to extract useful information from the data. The dataset was collected by Orange Labs for the region of Ivory Coast. The data in the dataset contains the \emph{Call Detail Records} of the calls made by users in the region of Ivory Coast \cite{Blondel2012}. The dataset contains both coarse grain mobility information as well as fine grain mobility information. The coarse grain mobility dataset is generated using subprefectures in Ivory Coast while fine grain dataset is generated using antenna locations. The dataset we use in this section is the course grain dataset and has the following structure. It contains the central location of the subprefecture area in the longitude and latitude format. The dataset contains information of 500,000 users collected over 5 months from 5th December 2011 until 28th April 2012. Each tuple in dataset contains \emph{subprefectureID} from where the call was made, \emph{time} when the call was made and the \emph{userID} of the users who made the call. Using aforementioned information we can estimate mobility pattern of a user. We decided mainly to use the coarse grain dataset, instead of the fine grain one, because it contains a larger user base.

Our first analysis (cf. Fig. \ref{fig:distributiona}) of the dataset, in accordance with the previous analysis of \cite{gonzálezunderstanding2008, brockmannthe2006, colizzaepidemic2008}, show that the flow distribution of the users movements display strong heterogeneities. The analysis highlights the fact that the flow of users toward some subprefectures is unevenly distributed (for example, the capital city subprefecture attracts more people than others. Cf. Fig. \ref{fig:graph} where the graph depicts the transition structure between subprefectures in Ivory Coast. Here, a node is plotted according to their geographical coordinates and their size is proportional to the number of incoming transitions.). Secondly, we show that the choice of subprefecture as the base level for our metapopulation model introduces only a small bias in the distribution of the jump length (cf. Fig. \ref{fig:distributionb}. A summary of various statistics is shown in Table \ref{table1}). Note that, in Fig. \ref{fig:distributionb}, the dashed line represents the distribution of the distances between subprefectures, the dots represent the distribution of jump lengths made by users when they make transitions between two subprefectures and the dashed line represents the power law fit of jump length distribution with parameter $\alpha = 2.51 \pm 0.0032$. Further, if we compare the distribution of jump length based on the subprefectures with the distribution of jump length made based on the position the antennas (cf. Fig. \ref{fig:distributionb}), the former distribution is only slightly truncated at the head and at the tail as compared to the original distribution (cf. Fig. \ref{fig:distributionb}). Note that the flow distribution at both subprefecture level and at antenna level is calculated using complete dataset. 

Further, as an example, we track a user $u$ (\emph{userID}$=297412$ in the dataset) (cf. Fig. \ref{fig:mobilityeuclid}). Here, we show subprefectures reached by the user $297412$. In the fig. \ref{fig:mobilityeuclid} the width of the line represents the number of transitions made by the user between a pair of subprefecture and each dot represents the subprefecture location. The size of the subprefecture shows the time a user stays at a given subprefecture. The yellow subprefecture marks the home subprefecture of the user (cf. Def. \ref{defhome}).

\begin{table}[!t]
\centering
\begin{tabular}{|c|l|l|l|l|l|}
\hline
{\footnotesize Rank} & {\footnotesize Subprefecture} & {\footnotesize In-flow} & {\footnotesize Betweenness} & {\footnotesize Population} & {\footnotesize Area} \\
& & & {\footnotesize Centrality} & {\footnotesize Density (approx)} & {\footnotesize (approx)}\\
\hline
1 & Abidjan & 0.1161 & 0.069256921 & 423.95 & 586.95\\
2 & Bingerville & 0.0371 & 0.010260404 & 423.95 & 257.66\\
3 & Soubre & 0.0323 & 0.017103843 & 54.39 & 1256.07\\
4 & Meagui & 0.0249 & 0.010316916 & 54.39 & 3491.10\\
5 & Anyama & 0.0245 & 0.030273783 & 423.95 & 695.02\\
6 & Yamoussoukro & 0.0240 & 0.043985509 & 67.19 & 1330.73\\
7 & Songon & 0.0216 & 0.022672720 & 423.95 & 623.31\\
8 & Grand-Bassam & 0.0180 & 0.011894960 & 52.56 & 124.98\\
9 & Bonoua & 0.0156 & 0.010142240 & 52.56 & 802.70\\
10 & Tiassale & 0.0156 & 0.030023932 & 57.24 & 2807.50\\
\hline
\end{tabular}
\caption{Top 10 subprefectures ranked by their in-flow (per-capita). The table also lists the betweenness centrality of the subprefectures normalized by $N(N-1)$ (for directed graph) where N is the number of nodes in the graph, the respective population density ($people/km^2$) and area ($km^2$).}
\label{table1}
\end{table}

%
%
\subsection{Interpreting mobility information}\label{subsec:Data set_2_and_3}

Let $C$ be the set of subprefectures. Let $P \in \mathbb{R}^{|C| \times |C|}$ be a weighted matrix where $P_{ij}\geq 0$ represents the number of transition between subprefecture $i$ and $j$. Note that a transition here represents the number of times users in subprefecture $i$ moved to subprefecture $j$. Also note that $P_{ii}=0$ and $P$ is an aggregated matrix of transitions made by all the users tracked in the $D4D$ dataset.

\newtheorem{definition}{Definition}

%
%

\begin{definition}\label{defmpv}
	Let $\nu\in \mathbb{R}^{|C|\times|C|}$ be a row stochastic matrix where the element $\nu_{ij}$ estimates the conditional probability of moving from subprefecture $i$ to subprefecture $j$ given that it has moved out of subprefecture $i$. We define $\nu_{ij}$ $\forall i \neq j$ as
	\begin{equation}\label{eq:mpv}
		\displaystyle \nu_{ij}=\frac{P_{ij}}{\sum_{\forall k \in C}P_{ik}}.
	\end{equation}
	where $\displaystyle \sum_{\forall j\in C}\nu_{ij}=1$ and $\nu_{ii}=0$. Note that $\nu_{ij} \in \mathbb{R}^+$.
\end{definition}

%
%
\begin{definition}\label{defmprv}
Let $\sigma\in \mathbb{R}^{|C|}$ where the element $\sigma_{i}$ represents the rate of moving out of subprefecture $i$. We define $\sigma_{i}$ $\forall i \in C$ as
\begin{equation}\label{eq:mprv}
\displaystyle \sigma_i=\frac{\sum_{\forall k \in C}P_{ik}}{Time}
\end{equation}
where $Time=150\times24\times60$ minutes, the total duration in minutes over which SET3 was generated. Note that $\sigma_{i} \in \mathbb{R}^+$. Also note that when $\sigma_{i}\rightarrow \infty$ it means that the user does not stay in subprefecture $i$ and the movement out of the subprefecture occurs instantaneously. However, when $\sigma_{i}\rightarrow 0$, it means that a user stays permanently in subprefecture $i$.
\end{definition}

We now formulate the rate at which the user returns to its home subprefecture. For this we first define home subprefecture as follows.

%
%
\begin{definition}\label{defhome}
Let $U_{u} = (C^{1}_{t_1}, C^{2}_{t_2}, \dots, C^{n}_{t_n})$, be a list of subprefectures visited by user $u$ and let  $t_{n} - t_{n-1}$ be the time spent in subprefecture $C^{n-1}$. Let  $M^{u}_j$ be the total time spent by user $u$ in subprefecture $j$. The home subprefecture $k$, of the user $u$ is the subprefecture that satisfies
\[k = \operatorname*{arg\,max}\limits_{\forall j \in C} M^{u}_j.\]
\end{definition}

%
%
Let $P^r \in \mathbb{R}^{|C|\times |C|}$ be an aggregated weighted matrix where $P^{r}_{ji}\geq 0$ represents the number of times users returns to their home subprefecture $i$ from subprefecture $j$.

%
%
\begin{definition}\label{defmprrv}
Let $\zeta\in \mathbb{R}^{|C|\times|C|}$ where the element $\zeta_{ji}$ represents the rate of coming back to the home subprefecture $i$ from another subprefecture $j$. We define $\zeta_{ji}$ $\forall i \in C$ as
	\begin{equation}\label{eq:mrr}
		\displaystyle \zeta_{ji}=\frac{P^{r}_{ji}}{Time}.
	\end{equation}
where $Time=150\times24\times60$ minutes, the total duration in minutes over which SET3 was generated. Note that $\zeta_{ji}\in\mathbb{R}^+$. Also note that when $\zeta_{ji}\rightarrow \infty$, it means that the device returns to its home community $i$ instantaneously after visiting community $j$. However, when $\zeta_{ji}\rightarrow 0$, it means that a device never moves back to its home community.
\end{definition}

In the dataset a user is associated to one and only one subprefecture at a given time. Using this information we determine the number of users in each subprefecture $P^T \in \mathbb{R}^{|T|\times|C|}$ at a given \emph{time slot} $t \in T$ where $T$ is the set of time slots in the dataset SET3.
%
%
\begin{definition}\label{defdm}
The user density matrix, $\varrho\in\mathbb{R}^{|T|\times|C|}$, with elements $\varrho_{ti}$ is defined as
\begin{equation}\label{eq:dm}
\varrho_{ti}=\frac{P^{T}_{ti}}{A_{i}}
\end{equation}
where $A_{i}$ is the area of the subprefecture $i$.
\end{definition}

\begin{figure}
    \centering
    \begin{subfigure}[b]{0.35\textwidth}
        \includegraphics[width=\textwidth]{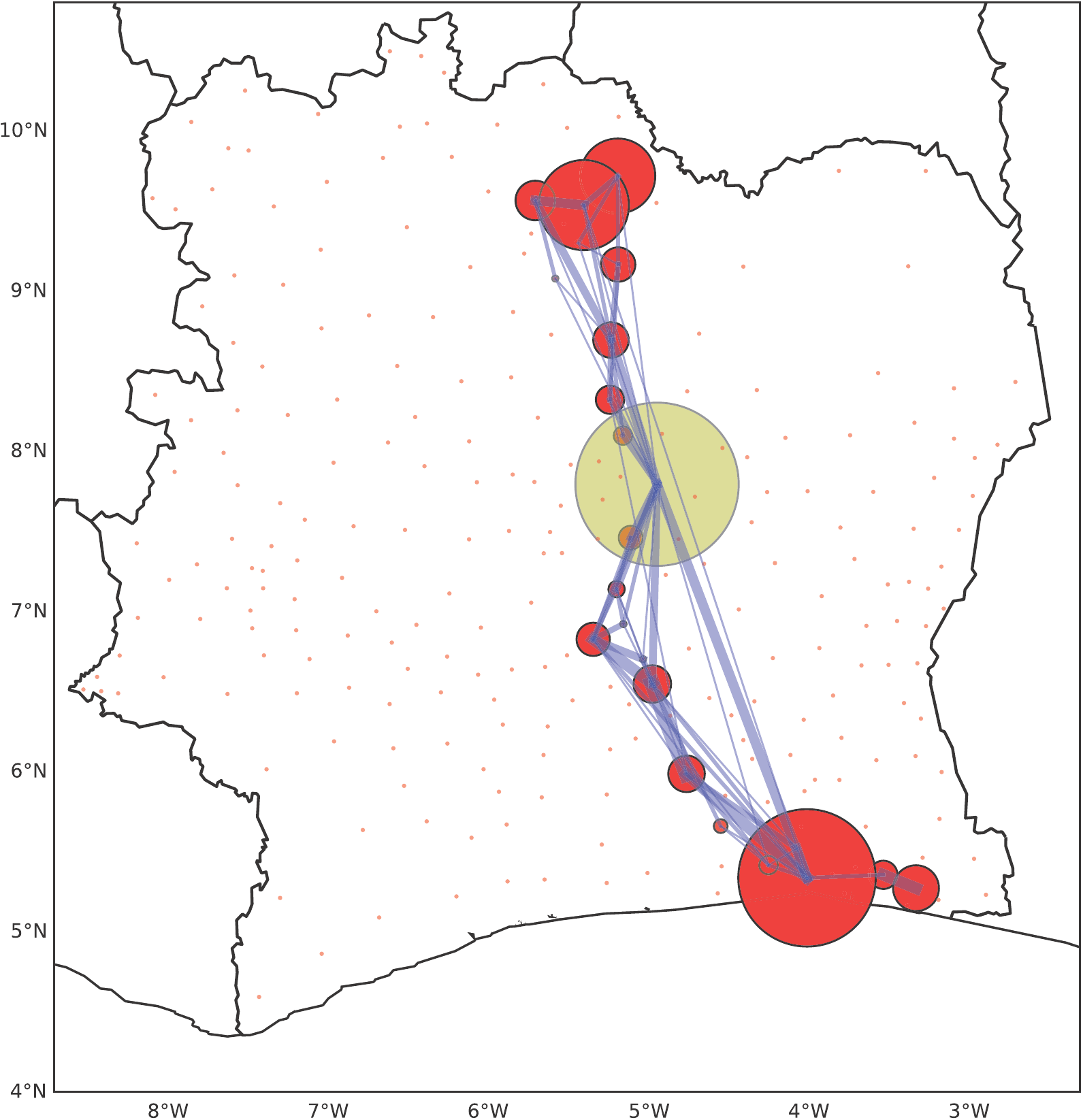}
	\caption{}
        \label{fig:mobilityeuclid}
    \end{subfigure}
    \begin{subfigure}[b]{0.55\textwidth}
        \includegraphics[width=\textwidth]{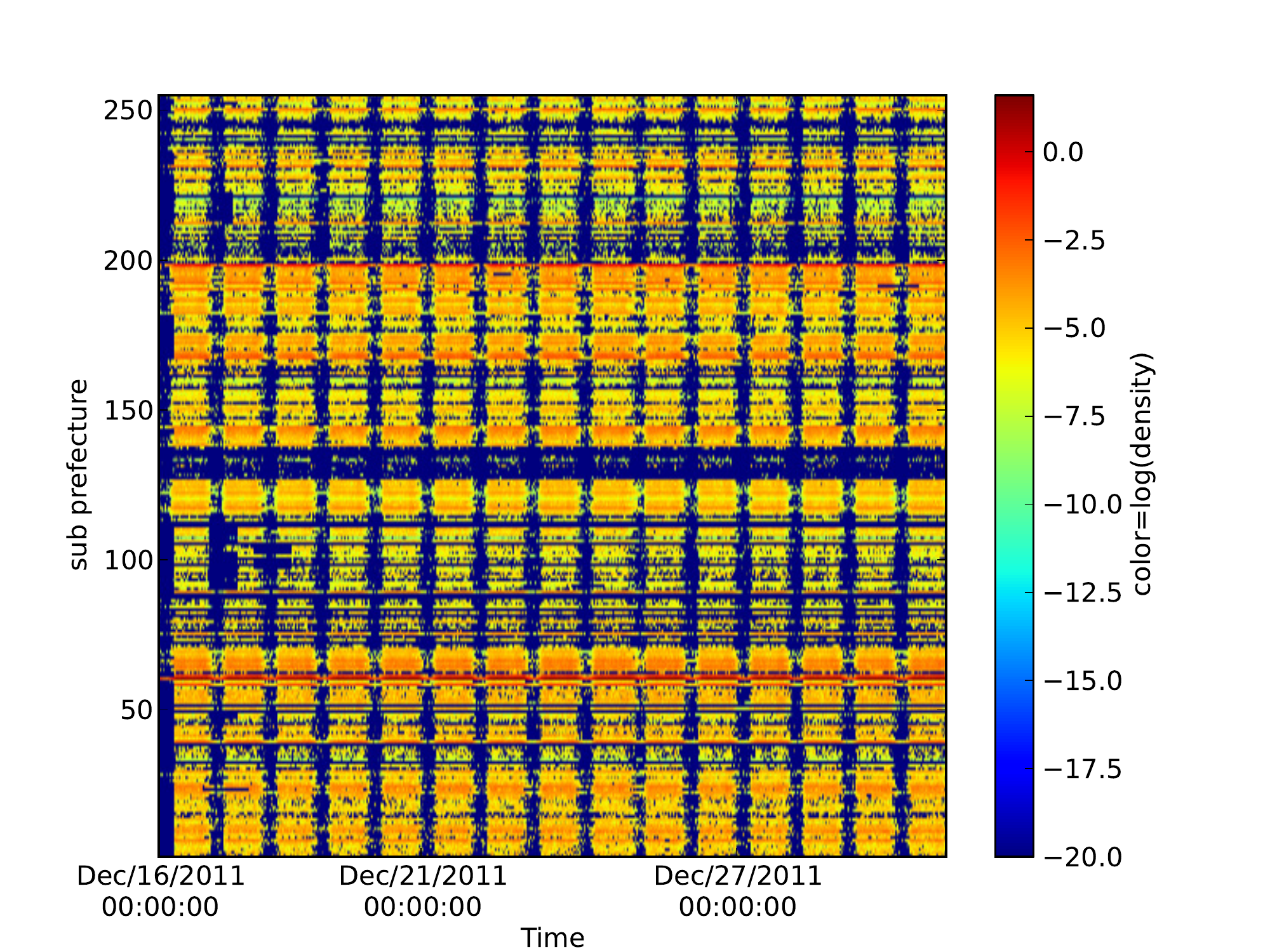}
	\caption{}
        \label{fig:densityuserdata}
    \end{subfigure}
    \caption{a) Trace of the user identified by number 297412, b) Density of the users in a subprefecture from 16th December 2011 until 31st December 2011 extracted from the dataset3. }
\end{figure}

In Fig. \ref{fig:densityuserdata} we provide a visual representation for the density of users on a log scale for the period from 16th December 2011 starting at 00:00:00 until 31st December 2011 ending at 23:59:00. The Fig. \ref{fig:densityuserdata} shows the variation of number of people in a subprefectures with respect to time. As the dataset is a $CDR$ where the information is logged only when a user makes a call, this variation relates to the fact that the number of calls made vary over time. Thereby suggesting that people tend to not communicate during certain period of time and thus not connect with the existing network at that time. In a $D2D$ perspective this change causes the network structure to vary over time and it thus provides us the motivation to incorporate the changes in the density through the concept of Latent States explained in Section \ref{subsec:latent}.

The information in the dataset is also used to infer the number of calls made in a subprefecture over time and also to analyze different properties related to the calls made. However, due to the scope of the paper we limit ourselves from providing these results.

We shall now describe our dissemination model using $\sigma$, $\nu$ and $\zeta$ in the next section.

%
%
\section{Model}\label{sec:model}

We represent our model in three layers. The bottom layer is the population layer where the population is spatially divided into communities. Note that we represent a person in the the population as a device. Above the population layer is the mobility layer where devices are allowed to jump from one community to another. Thus dissemination model is then applied on the top of the mobility model. The states in the dissemination model to which a device is associated to is either \emph{Susceptible} or \emph{Infected} or \emph{Recovered}. We denote these states as $S$, $I$ and $R$ respectively.

%
%
\subsection{Mobility Model}\label{subsec:C3_moblitymodel}
We consider a population of $N$ devices distributed among $|C|$ subprefectures according to the census data \cite{Web1}. Let $\rho_i = N_i/A_i$ be the density of devices in subprefecture $i$ where $N_i$ is the population in subprefecture $i$ and $\sum_{i\in C}N_i=N$. In a $D2D$ context a subprefecture is the area where the population of devices is well mixed and can contact other devices when they are in close proximity to each other. Further, as argued by Watts \emph{et al}., community structure is evident in the population in a realistic scenario \cite{Watts2005}. As each device is carried by humans this allows us to state that each device closely follows the mobility pattern of humans. The movement of a device or the jump of a device from a community $i$ to another community $j$ occurs with a rate $\sigma_i \nu_{ij}$ where $\sigma_i$ denotes the total rate out of community $i$ and $\nu_{ij}$ denotes the conditional probability of going to community $j$ given that there is a transition out of community $i$. We use $D4D$ SET3 as discussed in Section \ref{subsec:Data set_2_and_3} to determine $\sigma$ and $\nu$. It was shown that the jump also depends on the nature of the community \cite{Wang2012} and the activity pattern \cite{Liu2013}.

Further, we also consider rate of return to home community for each device \cite{Sattenspiel1995,Belik2011} and use $\zeta$ towards rate of return. Let $N_{ii}$ be the number of devices having home community as $i$ that are in community $i$ and $N_{ij}$ be the number of devices having home community as $i$ that are in community $j$.

\begin{figure}
    \centering
    \includegraphics[width=0.6\textwidth]{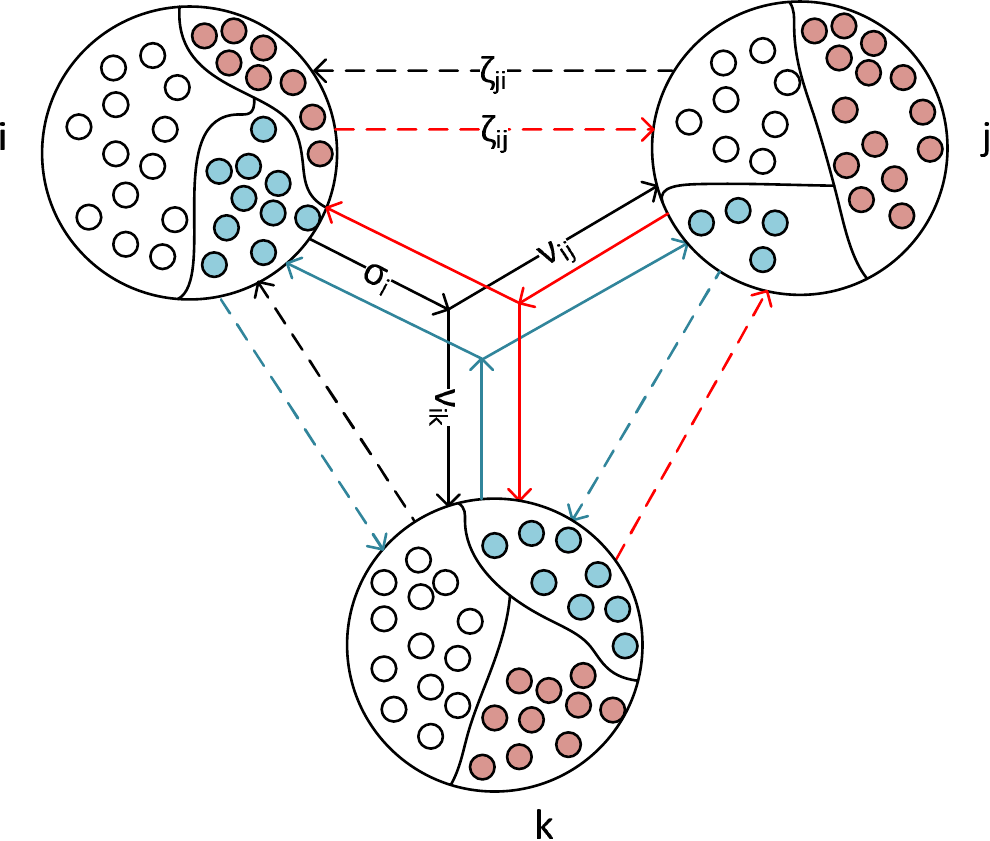}
    \caption{Mobility model with return rates.}
    \label{fig:mobilityVincent}
\end{figure}

Fig. \ref{fig:mobilityVincent} shows a mobility model with only three communities where movement from one community $i$ to another community $j$ is represented with an edge defined by $\sigma_{i}\nu_{ij}$. The dotted line shows the return rate from another community and is marked with $\zeta_{ji}$ and dotted line and rate of moving out of a community is given by $\sigma_i$. Note that each community $i$ in the above model consists of devices. Each community has associated number of devices. For different communities these devices are marked in different color (white for $i$, red for $j$, blue for $k$). The color of the lines also represents which community the devices are associated to. All other notations for movements are similarly defined. Moreover, this can be generalized to $|C|$ communities. This type of mobility model is a simple depiction of the inter-community movements with return rates.

Note that $\displaystyle \sum_{\forall i\in C}\sum_{\forall j\in C}N_{ij}=N$. Due to mobility, the number of the devices in a community $i$ changes. As represented in Fig. \ref{fig:mobilityVincent} the travel pattern leads to \cite{Sattenspiel1995} where the change in the number of devices in a community is defined as

\begin{eqnarray}\label{eq:NsteadyVincent}
    \left.\begin{matrix}\begin{aligned}
        \frac{dN_{ii}}{dt} &= \sum_{\forall j\in C, j\neq i} \zeta_{ji} N_{ij} - \sigma_{i}N_{ii}\\
        \forall j\in C, j\neq i, \frac{dN_{ij}}{dt} &= \sigma_{i}\nu_{ij} N_{ii} - \zeta_{ji} N_{ij}
    \end{aligned}\end{matrix}\right\}
\end{eqnarray}
The first term on the right hand side of the first equation in eq. \ref{eq:NsteadyVincent}, that is, $\displaystyle \sum_{\forall j\in C, j\neq i} \zeta_{ji} N_{ij}$, relates to the number of devices returning to their home community $i$ from other communities while the second term of the same equation, that is, $\sigma_{i}N_{ii}$, refers to devices that are moving out of community $i$ that have home community $i$. The second equation in eq. \ref{eq:NsteadyVincent} can also be similarly explained. Note that similar equations apply to all other communities. Also note that the total number of such equations is $|C| \times |C|$ and $\displaystyle N_i=N_{ii}+\sum_{j\in C,j\neq i}N_{ij}=\sum_{j\in C}N_{ij}$, $\displaystyle \sum_{j\in C}N_{i}=N$.

In order to visualize the dissemination process, we first identify the steady state of the mobility model. Note that there are two models, mobility model and dissemination model. For the mobility model, we look for the steady state limit solution. Let $N_{ii}^*$ and $N_{ij}^*$ be the steady state population of devices having home location as $i$ and are in community $i$ and $j$ respectively. Note that $N_{ii}^*$ is a limit and will not change over time. Then, we use the differential equations in order to derive the transient solution for the dissemination process where we study the change in the number of devices in $S$, $I$ and $R$ state.

Using (\ref{eq:NsteadyVincent}) we derive in (\ref{eq:NsteadyVincent2}) the number of devices per community at steady state \cite{Sattenspiel1995}.
\begin{eqnarray}\label{eq:NsteadyVincent2}
    \left.\begin{matrix}\begin{aligned}
        N_{ii}^{*} &= N_{i} \left( \frac{1}{1 + \sigma_{i} \sum_{\forall k \in C, k\neq i} \frac{\nu_{ik}}{\zeta_{ki}}} \right)\\
        \displaystyle \forall j\in C, j\neq i, N_{ij}^{*} &= N_{i} \left( \frac{\sigma_{i}\nu_{ij}}{ \zeta_{ji} \left(1 + \sigma_{i} \sum_{\forall k\in C, k\neq i} \frac{\nu_{ik}}{\zeta_{ki}} \right)} \right)
    \end{aligned}\end{matrix}\right\}
\end{eqnarray}

%
%
\subsection{Dissemination Model}\label{subsec:C3_dissmodel}
We now describe the dissemination model. Let each device in a community be in either of the three states $S$, $I$ and $R$. State $S$ means that the device does not have the message. State $I$ means that the device has the message and is transmitting it. State $R$ means that the device has the message and is not transmitting the message as the $TTL$ of the message is expired.

Let each device have an omnidirectional transmission range, $r$ such that $r\in \mathbb{R}^+$. We assume the interactions among the devices to occur under mean field approximation where the expected number of contacts per device is given by $\langle k_{i}\rangle$.
\begin{definition}\label{defavdeg}
The expected neighborhood size for a device in a community $i$ is defined as $\langle k_{i}\rangle=\rho_{i}^{*}\pi r^2$ where $\rho_{i}^{*}$ is the population density at the mobility steady state and $r$ the typical transmission range of a device.
\end{definition}

We assume that the devices in different communities have different $\beta$ represented as $\beta_{i}$. 

\begin{definition}\label{defbeta}
In a community $i$ the transition rate with which a message is received successfully is given by $\beta_i  = -\log(1-c_i)$ where $c_i$ is the probability of correct transmission given a contact between two devices. It follows that the average number of newly people infected in community $i$ per unit of time is $\langle k_{i}\rangle \frac{\beta_{i}}{N_{i}^{*}} \sum_{\forall j \in C}  S_{ii} I_{ji}$
\end{definition}

The variation in $\beta$ across different community is because of many factors like, the density of the population and willingness of other devices to accept the transmitted information. This idea is analogous to the epidemic spreading in different communities which happens at different rates. Thus, devices in each community have different contact rates.

\begin{definition}\label{defdelta}
$1/\delta_{i}$ is the expected time a device having a information will spread it to his neighborhood. In another word $\delta_{i}$ is the rate with which devices in community $i$ change their state from $I$ to $R$.
\end{definition}

\begin{figure}
    \centering
    \includegraphics[width=0.6\textwidth]{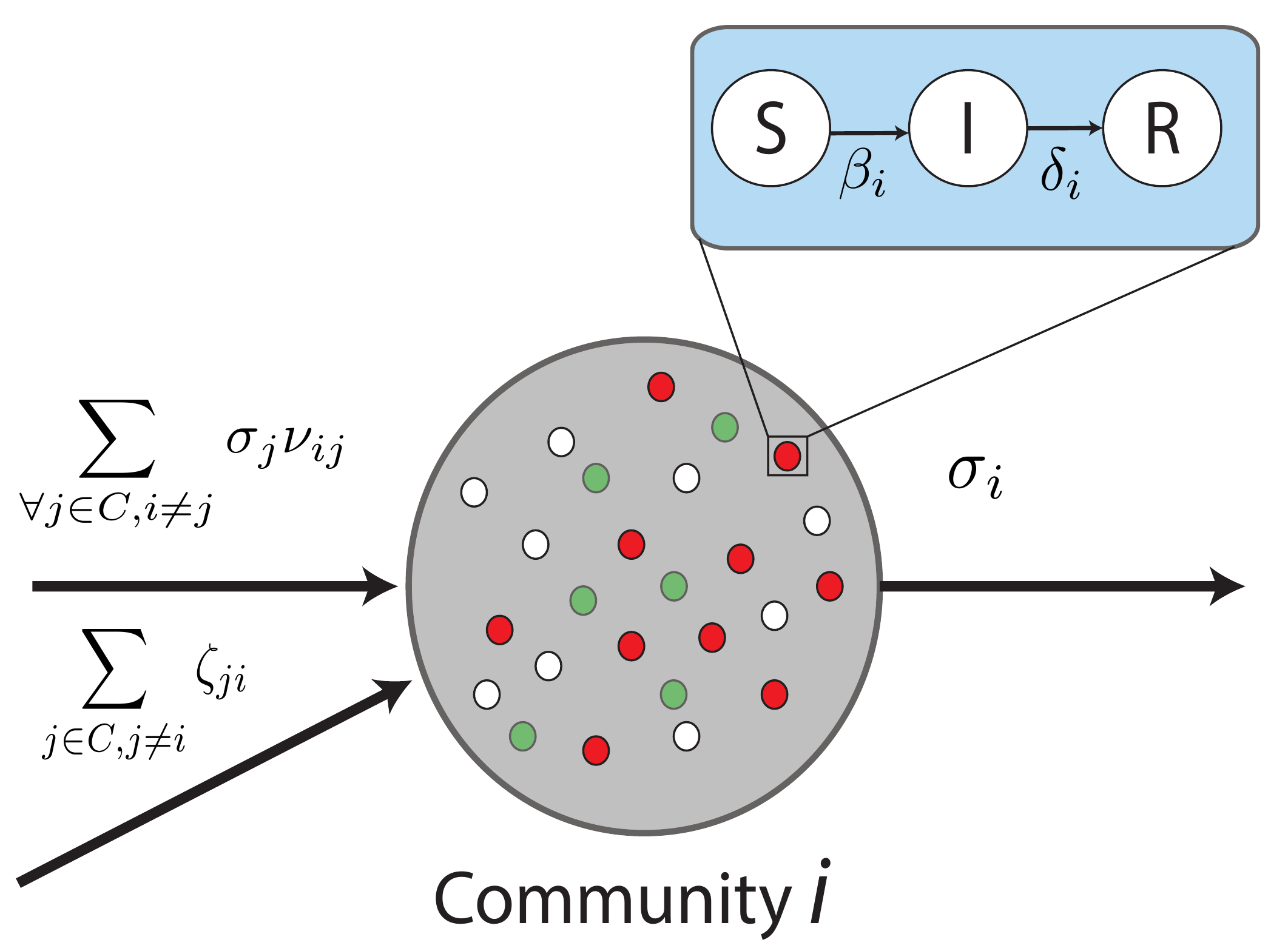}
    \caption{State diagram with states $S$, $I$ and $R$ with transition rates between states. The figure also shows the mobility transitions.}
    \label{fig:SIR1}
\end{figure}

In this scenario, the nodes that did not receive the public safety information yet (node in state $I$ inside community $i$), will receive the information at rate $\langle k_{i}\rangle \beta_i$ and devices that have already received the information will stop sending it at rate given by $\delta_{i}$ (cf. Fig \ref{fig:SIR1}). Note that, this process occurs independently in each community. Due to the energy constraints, a device can only transmit information for a given period of time after which the device stop transfer the information. This can be understood as the rate with which devices change their state from $I$ to $R$.

Similar to the explanation of $N_{ii}$ and $N_{ij}$, let $S_{ii}$, $I_{ii}$ and $R_{ii}$ be the number of devices in state $S$, $I$ and $R$ respectively, that have the home community as community $i$ and are in the community $i$ while $S_{ij}$, $I_{ij}$ and $R_{ij}$ be the number of devices in state $S$, $I$ and $R$ respectively, that have the home community as community $i$ and are in the community $j$. As we apply the dissemination model on top of the mobility model where we approximated the number of devices in each community by the steady state limit, $\displaystyle N_{i}^{*}=\sum_{\forall j\in C}N_{ij}^{*}=\sum_{\forall j\in C}\left[ \mathbf{E}[S_{ij}]+\mathbf{E}[I_{ij}]+\mathbf{E}[R_{ij}]\right]$.

In a community $i$, let initially at time $t=0$, all devices be in susceptible state. In order to study the dissemination process, let one community $i$ at time $t=0$ be the source of information, i.e., $S_{ii}=N_{ii}^{*}-\varepsilon$, $I_{ii}=\varepsilon$ and $\displaystyle \sum_{\forall j\in C}R_{ij}=0$ where $0<\varepsilon<N_{ii}$. Under mean field approximation, $\displaystyle \frac{1}{N_{i}^{*}}\sum_{\forall j \in C}S_{ii}I_{ji}$ provides the fraction of interactions between devices in state $S$ that have home community as $i$ and are in community $i$ with the devices in state $I$ that have home community $j$ and are in community $i$. As each device has a neighborhood $\langle k_{i}\rangle$ and the contact rate between devices is $\beta_i$, the number of devices in community $i$ that will change their state from $S$ to $I$ is given by $\displaystyle \beta_{i}\frac{\langle k_{i} \rangle}{N_{i}^{*}}\sum_{\forall j \in C}S_{ii}I_{ji}$. Let us consider the change in the number of devices in state $S$. Due to mobility, some devices in state $S$ will return from other communities to community $i$ while some devices in state $S$ will move from community $i$ to other communities. The number of devices moving from other communities to $i$ is given by $\displaystyle \sum_{\forall j\in C, j\neq i}\zeta_{ji}S_{ij}$ while those moving from $i$ to other communities is given by $ \sigma_{i}S_{ii}$. Thus the rate of change in the number of devices in state $S$ in community $i$ that have home community $i$ and are in community $i$ in $dt$ time is given by $\displaystyle -\beta_{i}\frac{\langle k_{i} \rangle}{N_{i}^{*}}\sum_{\forall j \in C}S_{ii}I_{ji}+\sum_{\forall j\in C, j\neq i}\zeta_{ji}S_{ij} - \sigma_{i}S_{ii}$. Using such interpretation and above explanation the rate of change in the number of devices in state $I$ and state $R$ can be formulated. The rate equations for the change in number of devices in state $S$, $I$ and $R$ are given as

\begin{eqnarray}\label{eq:N_SimpleSIRVincent1}
    \small
    \left.\begin{matrix}\begin{aligned}
        \frac{d S_{ii}}{dt}&=-\beta_{i}\frac{\langle k_{i} \rangle}{N_{i}^{*}}\sum_{\forall j \in C}S_{ii}I_{ji}+\sum_{\forall j\in C, j\neq i}\zeta_{ji}S_{ij}\\
        & - \sigma_{i}S_{ii}\\
        \forall j\in C,j\neq i, \frac{d S_{ij}}{dt}&=-\beta_{j}\frac{\langle k_{j} \rangle}{N_{j}^{*}}\sum_{\forall q \in C}S_{ij}I_{qj}+\sigma_{i}\nu_{ij}S_{ii}\\
        & -\zeta_{ji}S_{ij}
    \end{aligned}\end{matrix}\right\}
\end{eqnarray}
\begin{eqnarray}\label{eq:N_SimpleSIRVincent2}
    \small
    \left.\begin{matrix}\begin{aligned}
        \frac{d I_{ii}}{dt}&=\beta_{i}\frac{\langle k_{i} \rangle}{N_{i}^{*}}\sum_{\forall j \in C}S_{ii}I_{ji}+\sum_{\forall j \in C, j\neq i}\zeta_{ji}I_{ij}\\
        & - \sigma_{i}I_{ii} - \delta_{i}I_{ii}\\
        \forall j\in C,j\neq i, \frac{d I_{ij}}{dt}&=\beta_{j}\frac{\langle k_{j} \rangle}{N_{j}^{*}}\sum_{\forall q \in C}S_{ij}I_{qj}+\sigma_{i}\nu_{ij}I_{ii}\\
        & -\zeta_{ji}I_{ij} - \delta_{j}I_{ij}
    \end{aligned}\end{matrix}\right\}
\end{eqnarray}
\begin{eqnarray}\label{eq:N_SimpleSIRVincent3}
    \small
    \left.\begin{matrix}\begin{aligned}
        \frac{d R_{ii}}{dt}&=\delta_{i}I_{ii}+\sum_{\forall j \in C, j\neq i}\zeta_{ji}R_{ij} - \sigma_{i}R_{ii}\\
        \forall j\in C,j\neq i, \frac{d R_{ij}}{dt}&=\delta_{j}I_{ij}+ \sigma_{i}\nu_{ij}R_{ii} -\zeta_{ji}R_{ij}
    \end{aligned}\end{matrix}\right\}
\end{eqnarray}
Note that the total number of such equations in the system is $3\times |C| \times |C|$ and the system is easy to solve.

For a single population model (when number of communities=1), the outbreak of epidemic will be reached if the basic reproduction number $R_0$ is reached. Under the homogeneous mixing of the population the basic reproduction number is
$$R_0 = \frac{\langle k \rangle\beta}{\delta} > 1$$
However, this did not take into account the mobility of an infected person. In the case of the metapopulation model, we need to take into account two thresholds. First, the local epidemic threshold, $R_0 > 1$, within each subpopulation. Second, the global invasion threshold, $R_{*} > 1$, that defines the travel rate of individual. More details on the two aforementioned thresholds for metapopulation model is available in \cite{24418011}.

%
%
\subsection{Dynamic users model}\label{subsec:latent}

Irrespective of whether a device is in state $S$, $I$ or $R$, the capability of the device to transmit or receive also depends on whether the device is switched on or off. It is possible that a device has the information but has been switched off by its user. This hampers the transmission of the information from the device to other devices. Also, if a device is switched off, the device will not be able to receive the information from other devices. We call such state of a device as latent state and term them as $E_{S}$, $E_{I}$ and $E_{R}$ to represent latent states pertaining to each active state, $S$, $I$ and $R$ respectively. When a device is switched on, this will mark the transition in the state of the device from either $E_{S}$, $E_{I}$ or $E_{R}$ to $S$, $I$ or $R$ respectively. Further, to clarify the differences between the states, we list below the definitions of all the states

\begin{itemize}
\item \textbf{State $S$} - means that a device is ON and does not have received the message.
\item  \textbf{State $I$} - means that a device is ON, it has the message in his memory, the TTL of the message has not expired and the device is transmitting it.
\item \textbf{State $R$} - means that the device is ON, it has received the message and the device stops transmitting it as the TTL of the message has expired.
\item \textbf{State $E_S$} - means that a device is switched OFF and does not have the message.
\item \textbf{State $E_I$} - means that a device is switched OFF, it has the message in its memory, the TTL of the message has not expired but the device cannot transmit the message as it is switched off. Also, the device can restart sending the message once it returns to the state I.
\item \textbf{State $E_R$} - means that a device is OFF, it has received the message and the device stops transmitting it as the TTL of the message has expired.
\end{itemize}

\begin{definition}\label{defdelta}
We define all the following parameters in $\mathbb{R}^+$. Let the transition rate to change the state from $S$ to $E_{S}$ in a community $i$ be $\mu_{S_i}$. Let the transition rate to change the state from $I$ to $E_{I}$ in a community $i$ be $\mu_{I_i}$. Let the transition rate to change the state from $R$ to $E_{R}$ in a community $i$ be defined as $\mu_{R_i}$. The transition rate to change the state from $E_{S}$ to $S$ in a community $i$ is defined as $\alpha_{S_i}$. The transition rate to change the state from $E_{I}$ to $I$ in a community $i$ is defined as $\alpha_{I_i}$. The transition rate to change the state from $E_{R}$ to $R$ in a community $i$ is defined as $\alpha_{R_i}$.
\end{definition}

Further, a device in state $E_I$ can wake up and decide not to transmit information.
\begin{definition}\label{defgamma}
The transition rate to change the state from $E_{I}$ to $R$ in a community $i$ is defined as $\gamma_{i}$ such that $\gamma_{i}\in \mathbb{R}^+$.
\end{definition}

Note that due to addition of three new states, $N_{i}^{*}=\sum_{\forall j\in C}[\mathbf{E}[S_{ij}]+\mathbf{E}[I_{ij}]+\mathbf{E}[R_{ij}]+\mathbf{E}[E_{S_{ij}}]+\mathbf{E}[E_{I_{ij}}]+\mathbf{E}[E_{R_{ij}}]$. Similar to initial condition as in the previous Subsection \ref{subsec:C3_dissmodel}, we assume at time $t=0$ $S_{ii}=N_{ii}^{*}-\varepsilon$, $I_{ii}=\varepsilon$ and $R_{ii}=E_{S_{ii}}=E_{I_{ii}}=E_{R_{ii}}=0$, where $0<\varepsilon<N_{ii}$.
\begin{figure}
    \centering
    \includegraphics[width=0.6\textwidth]{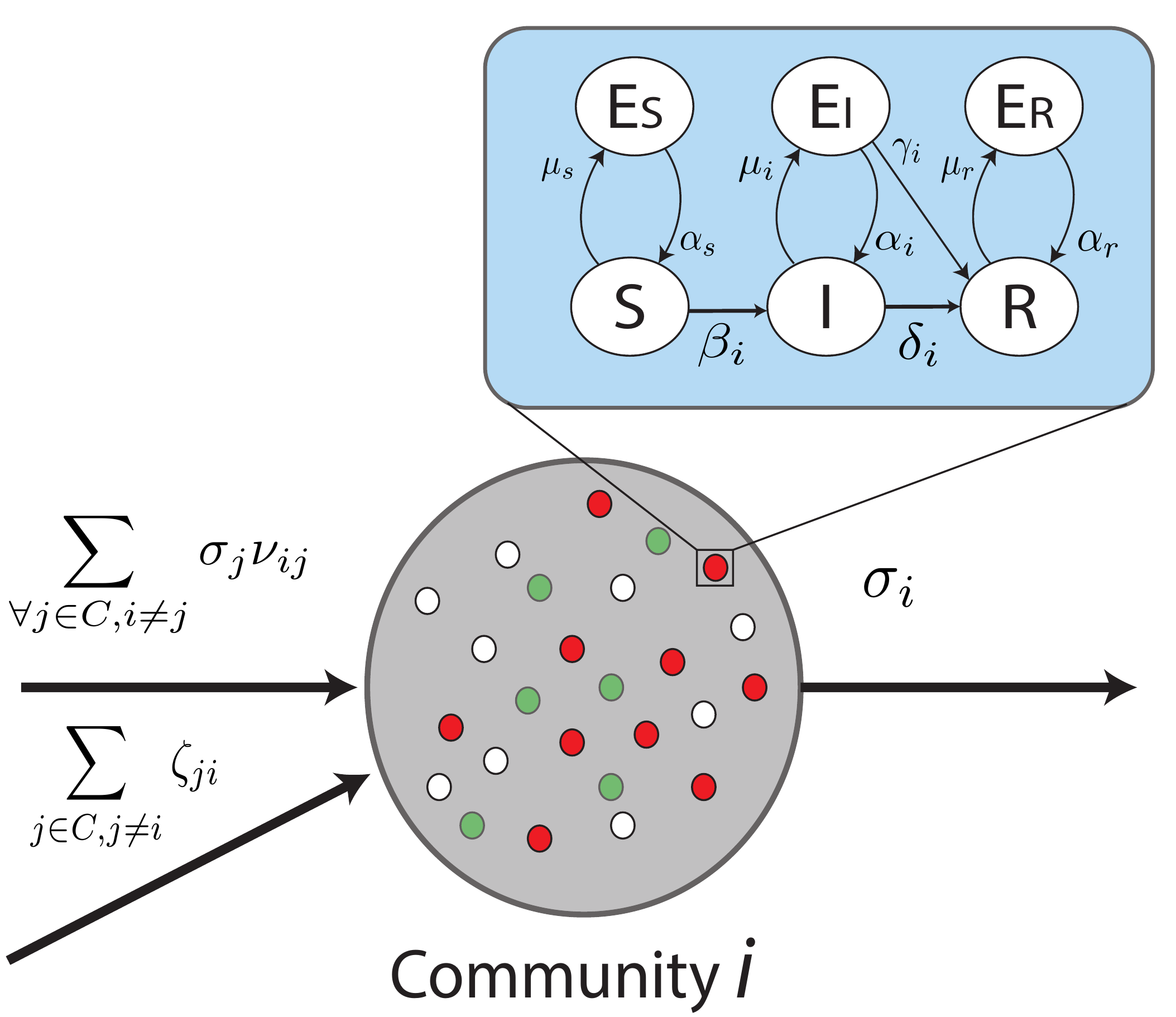}
    \caption{State diagram with states $S$, $I$ and $R$ and their latent states $E_{S}$, $E_{I}$ and $E_{R}$ respectively with transition rates between states. The figure also shows the mobility transitions.}
    \label{fig:spreadingmodel}
\end{figure}

Thus, the state diagram for this case is given by Fig. \ref{fig:spreadingmodel}. This addition of latent states leads us to modify the equations from (\ref{eq:N_SimpleSIRVincent1}) - (\ref{eq:N_SimpleSIRVincent3}) for a community $i$ to equations (\ref{eq:N_SimpleSIRbeamLatentVincent1}) -  (\ref{eq:N_SimpleSIRbeamLatentVincent6}). Note that, the total number of such equations in the system is $6\times |C| \times |C|$.

%
%
\begin{definition}\label{returnProba}
We borrow from \cite{polettoheterogeneous2012} the idea of an heterogeneous model to compute the dwell time for the initial population currently in subprefecture $i$ but having home location in subprefecture $j$. Let us assume that the return rate $\zeta_{ij}$ is $\zeta_{ij}^{-1} = \bar{\zeta}^{-1} \frac{d_{j}^{\chi}}{\langle d^{\chi}\rangle}$ where $\bar{\zeta}^{-1}$ is the average dwell time in our simulation and $d_i$ is the degree of connectivity of the subprefecture $i$ extracted from the transition matrix $\mathbf{P}$ between subprefectures, and $\langle d \rangle$ is the average degree of the subprefectures. The authors in \cite{polettoheterogeneous2012} introduced $\chi < 0$ in order to take in account the fact that low degree subprefectures are generally peripheral node in the graph. A trip from a given origin thus may take multiple steps to reach the final destination marked by peripheral locations, and thus a longer dwell time at the destination may then account for longer travel time to peripheral locations.
\end{definition}
This assumption fits the case of the Ivory Coast where the peripheral subprefectures (subprefectures with low degree centrality) have a very poor road infrastructure (only trails) and as a result they are more difficult to reach. Thus taking more time to reach them.

\begin{figure*}[htb]
\centering
\begin{minipage}{1.0\textwidth}
\small
\begin{flalign}
        &\left.\begin{matrix}\begin{aligned}\small\label{eq:N_SimpleSIRbeamLatentVincent1}
            \frac{d S_{ii}}{dt}&=-\beta_{i}\frac{\langle k_{i} \rangle}{N_{i}^{*}}\sum_{\forall j \in C}S_{ii}I_{ji}+\sum_{\forall j\in C, j\neq i}\zeta_{ji}S_{ij} - \sigma_{i}S_{ii}-\mu_{S_i}S_{ii} + \alpha_{S_i}E_{S_{ii}}\\
            \frac{d S_{ij}}{dt}&=-\beta_{j}\frac{\langle k_{j} \rangle}{N_{j}^{*}}\sum_{\forall q \in C}S_{ij}I_{qj}+\sigma_{i}\nu_{ij}S_{ii} -\zeta_{ji}S_{ij}-\mu_{S_j}S_{ij} + \alpha_{S_j}E_{S_{ij}}\\
        \end{aligned}\end{matrix}\right\}&
        \\
        &\left.\begin{matrix}\begin{aligned}\small\label{eq:N_SimpleSIRbeamLatentVincent2}
            \frac{d I_{ii}}{dt}&=\beta_{i}\frac{\langle k_{i} \rangle}{N_{i}^{*}}\sum_{\forall j \in C}S_{ii}I_{ji}+\sum_{\forall j \in C, j\neq i}\zeta_{ji}I_{ij} - \sigma_{i}I_{ii} - \delta_{i}I_{ii} -\mu_{I_i}I_{ii} + \alpha_{I_i}E_{I_{ii}}\\
            \frac{d I_{ij}}{dt}&=\beta_{j}\frac{\langle k_{j} \rangle}{N_{j}^{*}}\sum_{\forall q \in C}S_{ij}I_{qj}+\sigma_{i}\nu_{ij}I_{ii} -\zeta_{ji}I_{ij} - \delta_{j}I_{ij}-\mu_{I_j}I_{ij} + \alpha_{I_j} E_{I_{ij}}\\
        \end{aligned}\end{matrix}\right\}&
        \\
        &\left.\begin{matrix}\begin{aligned}\small\label{eq:N_SimpleSIRbeamLatentVincent3}
            \frac{d R_{ii}}{dt}&=\delta_{i}I_{ii}+\sum_{\forall j \in C, j\neq i}\zeta_{ji}R_{ij} - \sigma_{i}R_{ii}-\mu_{R_i}R_{ii} + \alpha_{R_i}E_{R_{ii}} + \gamma_{i}E_{I_{ii}}\\
            \frac{d R_{ij}}{dt}&=\delta_{j}I_{ij}+ \sigma_{i}\nu_{ij}R_{ii} -\zeta_{ji}R_{ij}-\mu_{R_j}R_{ij} + \alpha_{R_j} E_{R_{ij}}  + \gamma_{j}E_{I_{ij}}\\
        \end{aligned}\end{matrix}\right\}&
        \\
        &\left.\begin{matrix}\begin{aligned}\small\label{eq:N_SimpleSIRbeamLatentVincent4}
            \frac{d E_{S_{ii}}}{dt}&=\mu_{S_i}S_{ii} - \alpha_{S_i}E_{S_{ii}} + \sum_{\forall j \in C, j\neq i}\zeta_{ji}E_{S_{ij}}-\sigma_{i}E_{S_{ii}}\\
            \frac{d E_{S_{ij}}}{dt}&=\mu_{S_j}S_{ij} - \alpha_{S_j}E_{S_{ij}} + \sigma_{i}\nu_{ij}E_{S_{ii}}-\zeta_{ji}E_{S_{ij}}\\
        \end{aligned}\end{matrix}\right\}&
        \\
        &\left.\begin{matrix}\begin{aligned}\small\label{eq:N_SimpleSIRbeamLatentVincent5}
            \frac{d E_{I_{ii}}}{dt}&=\mu_{I_i}I_{ii} - \alpha_{I_i}E_{I_{ii}} + \sum_{\forall j \in C, j\neq i}\zeta_{ji}E_{I_{ij}}-\sigma_{i}E_{I_{ii}} - \gamma_{i}E_{I_{ii}}\\
            \frac{d E_{I_{ij}}}{dt}&=\mu_{I_j}I_{ij} - \alpha_{I_j}E_{I_{ij}} + \sigma_{i}\nu_{ij}E_{I_{ii}}-\zeta_{ji}E_{I_{ij}} - \gamma_{j}E_{I_{ij}}\\
        \end{aligned}\end{matrix}\right\}&
        \\
        &\left.\begin{matrix}\begin{aligned}\small\label{eq:N_SimpleSIRbeamLatentVincent6}
            \frac{d E_{R_{ii}}}{dt}&=\mu_{R_i}R_{ii} - \alpha_{R_i}E_{R_{ii}} + \sum_{\forall j \in C, j\neq i}\zeta_{ji}E_{R_{ij}}-\sigma_{i}E_{R_{ii}} \\
            \frac{d E_{R_{ij}}}{dt}&=\mu_{R_j}R_{ij} - \alpha_{R_j}E_{R_{ij}} + \sigma_{i}\nu_{ij}E_{R_{ii}}-\zeta_{ji}E_{R_{ij}}\\
        \end{aligned}\end{matrix}\right\}&
\end{flalign}
\end{minipage}
\label{eq:N_SimpleSIRbeamLatentVincent}
\end{figure*}

%
%
\section{Results}\label{sec:simresult}

\begin{figure}
    \centering
    \includegraphics[width=\textwidth]{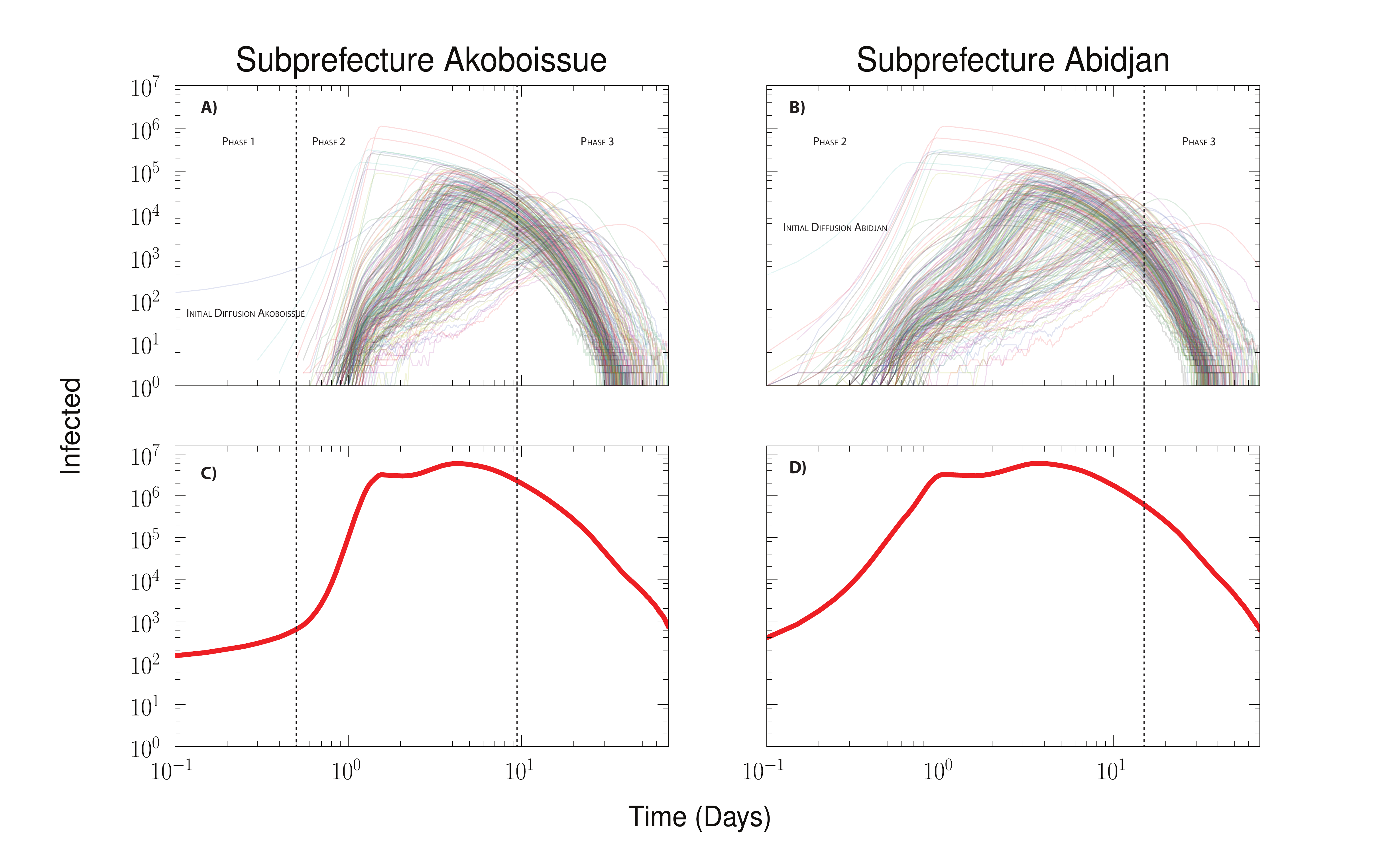}
    \caption{(\textbf{a}) and (\textbf{b}) Time evolution of the diffusion process across all subprefectures in the country for two different origins (sparsely populated and densely populated respectively) when latent states are not used. (\textbf{c}) and (\textbf{d}) show the overall fraction of infected devices at a given time for respective origins.}
    \label{fig:sirgillespi}
\end{figure}

We use the Gillespie ($\tau$-leap method \cite{Gillespie2001, Gillespie2006}) algorithm to solve the stochastic equations of the compartmental system of $255 \times 255 \times 6$ differential equations that describe the evolution of the system in each subprefecture (cf. \ref{AppendixA} for detail explanation of the simulation setup). We use census data of the Ivory Coast in the year 1998 \cite{Web1} to initialize the population of each subprefecture. Further, we compute surface area of each subprefecture through GIS mapping. Initially, we trigger a message inside a subprefecture \emph{Akoboissue} ($I_{Akoboissue}(0)=\varepsilon$ and $\forall i \neq Akoboissue$ $I_i=0$ where $\varepsilon=0.0052$, i.e., $\varepsilon=\frac{100}{18959}$) in one simulation and in \emph{Abidjan} in another ($I_{Abidjan}(0)=\varepsilon$ and $\forall i \neq Abidjan$ $I_i=0$ where $\varepsilon=4e-4$, i.e., $\varepsilon=\frac{100}{248842}$). We define the transmission radius of a device to be  $r=100m$. Through the evolution of overall percentage of devices in either of the three states, we show the effect on the dissemination of the public warning message across the population.

We first display as supplementary material a movie \cite{Agarwal2013aa} that shows the diffusion process in Ivory Coast. We observe that the diffusion initially takes place in the subprefecture \emph{Akoboissue} of the country. The information then spreads to major cities of Ivory Coast (as suggested by the mobility model cf. Fig. \ref{fig:graph}) through subprefecture \emph{Abidjan} (the economic capital), \emph{Bouake} (the second largest city), \emph{Yamoussoukro} (Political Capital) and \emph{Soubre}. The information then spreads slowly to other cities in the Ivory Coast. Also note that, the south east side of the country is well known to be mostly a cocoa producing agricultural region \cite{Web2}. It is also note worthy that the cocoa accounts for 25\% of the economic GDP of the country. \emph{Abidjan} being one major economic city in south east of Ivory Coast, accounts for high mobility of people and is well connected, thus also resulting into initial diffusion. We also note that the diffusion of the information first spreads to south eastern part of the country, then to south western. It then takes a long time to spread over the northern part of the country. The authors of \cite{Clio2013}, who worked on the same dataset, brought out the fact that one factor that the northern part of the country is less diffusive is due to the consequence of socio-economic disparity inside the country. They also brought out that this disparity can be a consequence of the fact that the northern part of the country is still relatively ``\textit{disconnected from the main economic and political center of Ivory Coast}". Further, it should also be noted that the northern part of the country is mostly \emph{Savanna} land with relatively less population.

In Fig. \ref{fig:sirgillespi}, we investigate the sensibility of our model to the placement of the initial diffusion. Through Fig. \ref{fig:sirgillespi} we show the time evolution of the diffusion process across the country (in all 255 subprefectures) using the model without latent states for two different origins \emph{Akoboissue} and \emph{Abidjan}. Fig. \ref{fig:sirgillespi}a and Fig. \ref{fig:sirgillespi}c show the case when the origin of the diffusion is located in a sparsely populated area (Akoboissue) while Fig. \ref{fig:sirgillespi}b and Fig. \ref{fig:sirgillespi}d show the case when the diffusion process originates from a dense area (the capital of the country - Abidjan). Through Fig. \ref{fig:sirgillespi}a and Fig. \ref{fig:sirgillespi}c we observe three distinct phases, phase 1: the messages are spreading only inside the initially infected subprefecture, phase 2: most of the subprefectures get infected, phase 3: the overall number of infected devices starts to decrease, however the infection continues in the peripheral subprefectures. Here it is observed that the diffusion reaches the major city of the country one day after the initial spreading (Phase 1), and further diffuses into the rest of the country in the time interval between 2 to 10 days (Phase 2). Finally, it reaches the northern part of the country after 10 days (Phase 3). However in contrast, via Fig. \ref{fig:sirgillespi}b and Fig. \ref{fig:sirgillespi}d, it is noted that when the diffusion is initiated in a denser area, the diffusion process progresses quickly throughout the country. Here phase 1 is almost non existent and the infection spreads to other neighboring subprefectures rapidly. Further, Fig. \ref{fig:sirgillespi} also support that the subprefectures in the north receive messages during the phase 3 (long time after the densely connected subprefecture of the south received the message). From the Fig. \ref{fig:sirgillespi}d we also observe that the first peak of infection takes place just less than one day after the initial diffusion started as compare to day and half in Fig. \ref{fig:sirgillespi}c.
\begin{figure}
    \centering
    \includegraphics[width=\textwidth]{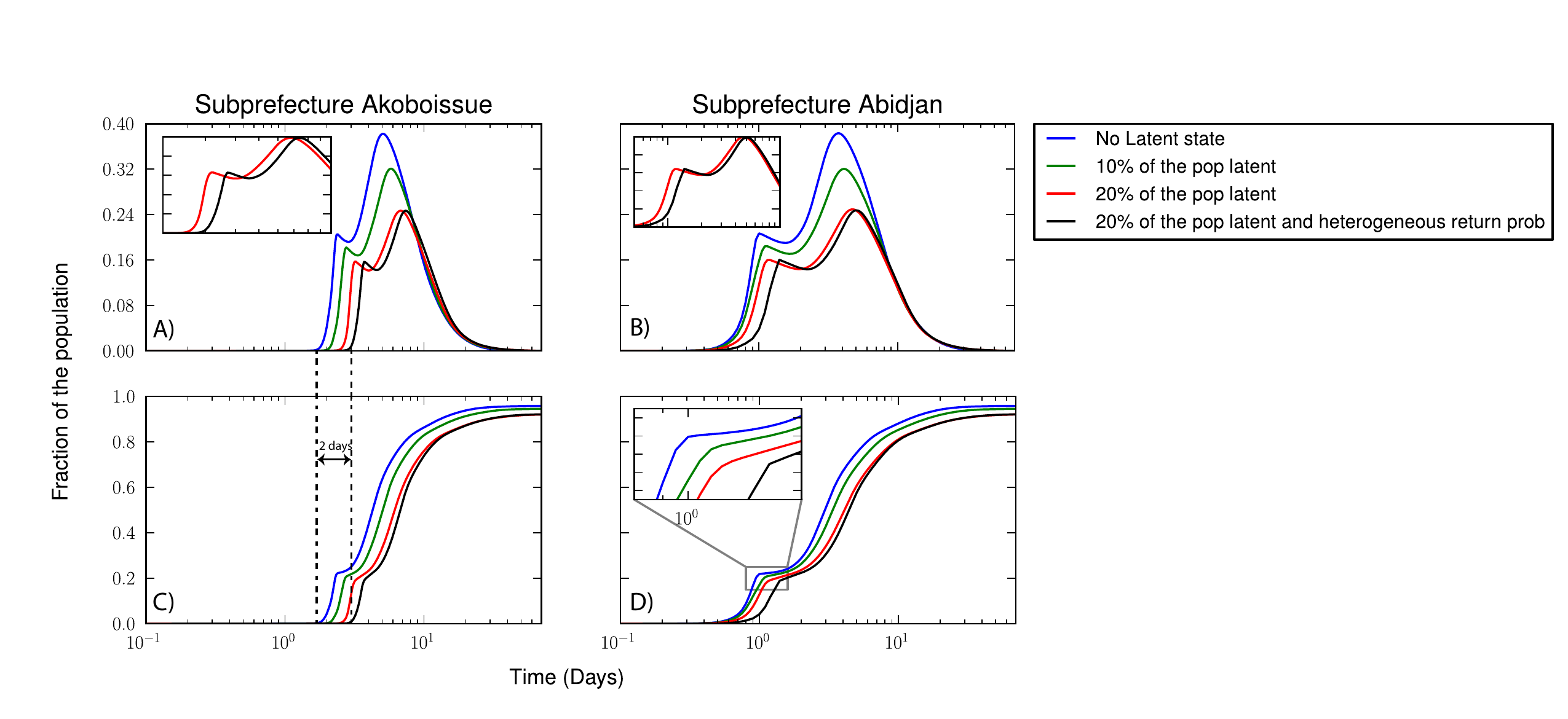}
    \caption{Time evolution of the number of infective device in Ivory Coast with different percentage of population in latent state. (\textbf{a}) when the diffusion starts from a sparsely populated area, (\textbf{b})  when the diffusion starts from a dense area, (\textbf{c}) Cumulative fraction of the population infected the initial infection is initiated from a sparsely populated area, (\textbf{d}) Cumulative fraction of the population infected when the initial infection is initiated from a densely populated area.}
    \label{fig:sirvssirlatent}
\end{figure}

It is also important to note that the diffusion process is enhanced if initial broadcast of the message started in the subprefecture that had high betweenness centrality. In this case the diffusion of the message is more quickly to a large portion of other subprefectures since its average distance to other subprefectures is relatively small (cf. Fig \ref{fig:sirgillespi}c and Fig. \ref{fig:sirgillespi}d). In the Fig. \ref{fig:sirgillespi}b and Fig. \ref{fig:sirgillespi}d the initial diffusion takes place inside the subprefecture of Abidjan which is both the most connected and also has the highest betweenness centrality (Table \ref{table1}).  

Finally from the Fig. \ref{fig:sirvssirlatent} we show the time evolution of the number of infective device in Ivory Coast when the latent state concept was added. Here, we temporally disable a given fraction of the population from being able to receive or transmitting message. In the Fig. \ref{fig:sirvssirlatent}, the blue curve refers to the model without latent state. The green and the red curves refer to the case when 10\% and 20\% devices are in latent state respectively. While the black curve reflects the evolution when 20\% of the population is latent and heterogeneous return probability (cf. Def. \ref{returnProba}) is used. Here, for heterogeneous case, we assume that the dwell time of people has an exponential distribution with a mean of $\frac{1}{2}$ day. The salient fact about this simulation is that removing temporally a fraction of the population from being active has for obvious consequence of delaying the peak of infection (cf. Fig. \ref{fig:sirvssirlatent}a and Fig. \ref{fig:sirvssirlatent}c). However, this effect is mitigated in the second case when we initiate the diffusion process from Abidjan (cf. Fig. \ref{fig:sirvssirlatent}b and \ref{fig:sirvssirlatent}d). Due to its strategic position inside the graph, in Abidjan (cf. Fig. \ref{fig:distribution}c) (highest centrality, highest connectivity, denser area cf. Table \ref{table1}) the removal of small fraction of the active population has almost no effect on the speed of the dissemination process (however the height of peak of infection remains attenuated as compared to the no latent case). This stems from the fact that in the case when the initial diffusion takes place in \textit{Akoboissue}, at the initial state the rate of migration (which is very low) of people carrying the message to neighboring (neighbors based on the flow of people, cf. Fig. \ref{fig:graph}) subprefectures is slowed down by people in inactive states (latent states), therefore it takes more time for neighboring prefecture to reach the critical threshold that enables them to sustain the diffusion process themselves.

%
%
\section{Discussion and Conclusion}\label{sec:conclusion}

In this paper, we presented a large scale data driven model for information dissemination in an heterogeneous network of relations (metapopulation) between subpopulation. We used the data provided by the $D4D$ organizers to determine the movement probabilities for the movement of the devices from one community to another. To realize the information dissemination process we used $SIR$ type epidemic model with addition of latent states $E_{S}$, $E_{I}$ and $E_{R}$ introducing dynamic users's behaviors. Thus, the paper presents two main contributions, first the introduction of latent states to account for variable density and second, how mobility in a network with community structure could be applied to achieve large scale dissemination in a dynamic $D2D$ based communication network.

\section{acknowledgement}\label{sec:acknowledgement}
The authors thank $D4D$ organizers for providing the data and gratefully acknowledge the benefits of useful discussions and comments from Francis Bu-Sung Lee.

\appendix
\section{Simulation Setup}\label{AppendixA}

\begin{figure*}
    \centering
    \includegraphics[width=\textwidth]{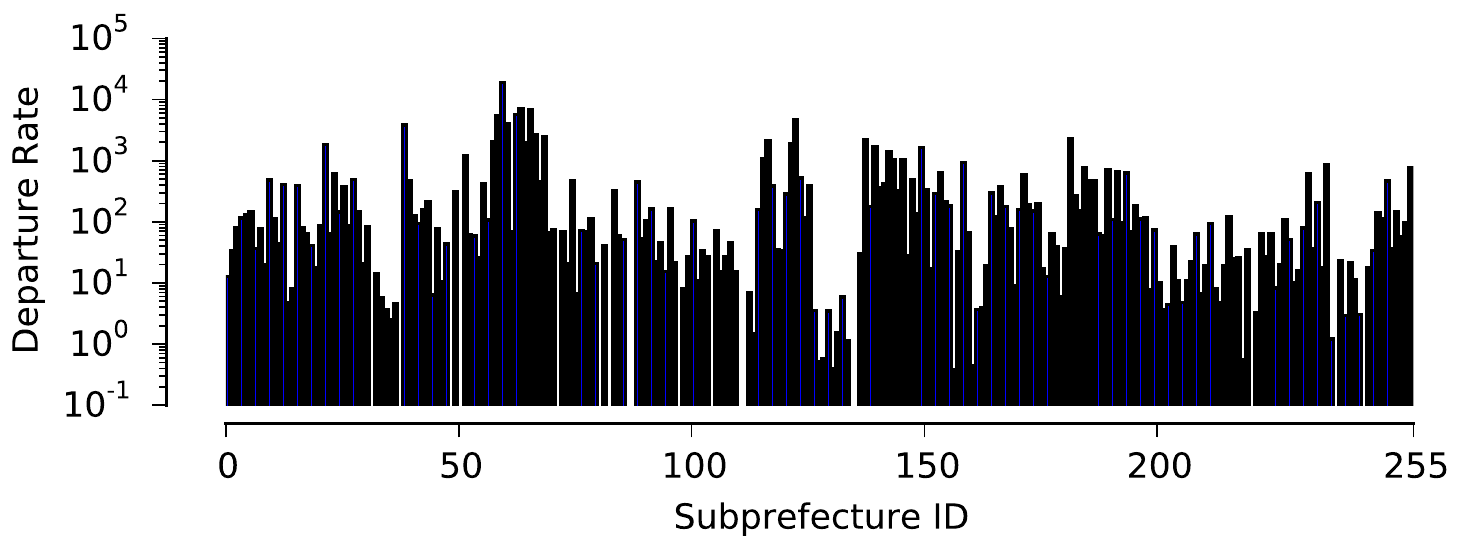}
    \caption{Departure rate from each subprefecture}
    \label{fig:departurerate}
\end{figure*}

During the simulation of the spreading process using the ``$\tau$-leap" method we modeled the departure rate as a Poisson process with specific mean for each subprefecture according to figure \ref{fig:departurerate}. The dwell time  in each latent states is modeled as an exponential time of half a day in the generic case and according to Definition \ref{returnProba} in the heterogeneous case. The rates $\mu_s, \mu_i, \mu_r$ are modeled as a Poisson process defined as function of the mean fraction of the population we wanted to put in latent states (either 10\% or 20\% of the total population). 

In our simulation, the diffusion process starts with a population repartition that is already in steady state. Nonetheless, one question remains, does the repartition of the population at steady state $D^{(2)}$, remains consistent with the initial repartition of the population $D^{(1)}$ (census data). In order to answer this question, we compare distributions quantitatively by calculating the symmetrized \textit{Kullback\hyp{}Leibler (KL)} divergence between the two partitions. The KL divergence between distribution $D^{(1)}=(D_i^{(1)})$ and $D^{(2)}=(D_i^{(2)})$ is defined as:

$$DIV_{KL}\left(D^{(1)}||D^{(2)}\right)= \frac{1}{2} \left(\sum_{i} D_i^{(1)}  \log \frac{D_i^{(1)}}{D_i^{(2)}} + \sum_{i} D_i^{(2)} \log⁡ \frac{D_i^{(2)}}{D_i^{(1)}} \right)$$

We find that the KL divergence of the order 0.001, meaning, there is only a marginal difference between the population distribution at initial state and at the steady state. To conclude our model of the population (model at steady state) does not deviate from original distribution of the population (census). However, it remains that the overall mobility of persons inside the country is highly skewed towards some prefectures (cf. Fig. \ref{fig:distribution}).

As a conclusion, the variance of the mobility process is bounded and does not evolve during the simulation, since we already computed its steady state before the simulation of the spreading process started. The simulation code used in this paper is available at the following address \cite{code}.

\bibliography{bibliography}
\end{document}